# Optoelectronically Active GaAs/GeSn-MQW/Ge Heterojunctions Created via Semiconductor Grafting

Jie Zhou[1,a)], Haibo Wang[2,a)], Yifu Guo[3,a)], Alireza Abrand[4,a)], Yiran Li[1], Yang Liu[1], Jiarui Gong[1], Po Rei Huang[5], Jianping Shen[1], Shengqiang Xu[2], Daniel Vincent[1], Samuel Haessly[1], Yi Lu[1], Munho Kim[6], Shui-Qing Yu[7], Parsian K. Mohseni[4,b)], Guo-En Chang[5], Zetian Mi[3], Kai Sun[8,b)], Xiao Gong[2,b)], Mikhail A Kats[1], and Zhenqiang Ma[1,b)]

*[1]Department of Electrical and Computer Engineering, University of Wisconsin-Madison, Madison, Wisconsin 53706, USA*

*[2]Department of Electrical and Computer Engineering, National University of Singapore 117576, Singapore*

*[3]Department of Electrical Engineering and Computer Science, University of Michigan, Ann Arbor, Michigan 48109, USA*

*[4]Department of Electrical and Microelectronic Engineering, Rochester Institute of Technology, Rochester, New York 14623, USA*

*[5]Department of Mechanical Engineering, and Advanced Institute of Manufacturing with High-tech Innovations, National Chung Cheng University, Chiayi County 62102, Taiwan*

*[6]School of Electrical and Electronic Engineering, Nanyang Technological University, 50 Nanyang Avenue, Singapore 639798, Singapore*

*[7]Department of Electrical Engineering, University of Arkansas, Fayetteville, Arkansas 72701, USA*

*[8]Department of Materials Science and Engineering, University of Michigan, Ann Arbor, MI 48109, USA*

[a)]  These authors contribute equally to this work.

[b)]  Author to whom correspondence should be addressed. Electronic mail: mazq@engr.wisc.edu; or mkats@wisc.edu

## Abstract

Traditionally, advancements in semiconductor devices have been driven by lattice-matched heterojunctions with tailored band alignments through heteroepitaxy techniques. However, there is significant interest in expanding the capabilities of heterojunction devices, in particular utilizing extreme lattice mismatches. We demonstrate the manipulation of device behaviors and performance enhancement achievable through a lattice-mismatched, single-crystalline GaAs/GeSn-multi-quantum well (MQW)/Ge *n-i-p* heterojunction by employing advanced semiconductor grafting technology. This approach allows for synergistic tailoring of optical and electronic properties within the structure. The grafted GaAs/GeSn-MQW/Ge heterojunction was comprehensively characterized. Detailed STEM studies confirm that the GaAs/GeSn-MQW heterointerface is atomically clean, without significant interfacial interdiffusion. A photodiode was fabricated based on this heterojunction. With engineered band alignment and optical field distribution, the grafted GaAs/GeSn-MQW/Ge *n-i-p* photodiode achieved outstanding performance: a record-low dark current density of $1.22 \times 10^{-7}$ A/cm$^2$, an extended spectral response from ~0.5 to 2 $\mu$m, and improved photoresponsivity of $R_{\text{VIS}}$ of 0.85 A/W and $R_{\text{NIR}}$ of 0.40 A/W at 520 and 1570 nm, respectively. The dark current density is at least 5 orders of magnitude lower than state-of-the-art GeSn photodiodes. The photoresponsivity demonstrates an approximately sevenfold enhancement in the VIS range and a threefold improvement in the NIR range compared to the reference epitaxial photodiode. This work presents a unique strategy for constructing lattice-mismatched semiconductor heterojunction devices. More importantly, the implications transcend the current GaAs/GeSn-MQW/Ge example, offering potential applications in other material systems and freeing device design from the stringent lattice-matching constraints of conventional heteroepitaxy.

**Introduction**

Semiconductor heterojunctions are a workhorse in modern information and communication technologies. Optoelectronic and electronic devices, such as double-heterostructure-based light-emitting diodes (LEDs), lasers, photodetectors, high-electron-mobility transistors (HEMTs), and bipolar junction transistors (BJTs) have facilitated the development of advanced technologies by harnessing the unique properties offered by heterojunctions. Fundamentally, a heterojunction is formed by interfacing two semiconductors with different bandgaps[1–4]. The resulting bandgap transition across the junction interface leads to an energy band discontinuity, also known as the "band offset". This band offset serves as either a potential barrier to confine the carriers, or an accelerator to promote carrier transport, depending on the device. As compared to homojunctions made from isotype materials, heterojunctions provide an additional dimension for manipulating device properties, including carrier transport, recombination, photon generation and detection.

Despite the extensive exploration and continuous potential of semiconductor heterojunctions, new challenges are still being uncovered. For instance, lattice mismatch[5,6] often causes interface imperfections like internal strain, defects, and dislocations, eventually contributing to a deterioration in the device performance.

This issue is particularly relevant for material systems like germanium-tin (GeSn), a promising narrow-bandgap semiconductor with direct bandgap from near to mid-infrared optoelectronics tunable with varying Sn compositions[7] and CMOS compatibility for low-cost mass production[8]. However, epitaxy of GeSn with high Sn content and the heteroepitaxy of GeSn heterostructures remain challenging[9]. As the difference in Sn content between adjacent GeSn layers increases, the lattice mismatch grows, leading to a higher density of defects and dislocations at the heterointerface, degrading the junction properties.

Additionally, the limited tunability of electronic and optical properties of epitaxial GeSn heterojunctions devices pose further challenges[10]. Typical GeSn heterojunctions, such as Ge/GeSn/SiGeSn with varying Sn or Si content, exhibit relatively small band offsets between the conduction and valence bands. Moreover, the small optical refractive index difference between Ge and GeSn restricts the modulation of the light distribution within GeSn heterostructures. These factors significantly limit the performance of GeSn optoelectronic devices created using conventional epitaxy. For example, previous GeSn-based photodiodes[11–25] generally exhibit mediocre rectifying characteristics, with relatively low rectification ratio and elevated dark-current density due to the small band offsets between neighboring Ge/GeSn layers.

Semiconductor grafting has emerged as an innovative solution to these issues[26-33]. This approach enables the formation of active heterojunctions between dissimilar semiconductors, irrespective of the difference of their lattice constants and thermal-expansion coefficients[34]. By utilizing semiconductor nanomembranes and ultrathin dielectric interlayers, the grafting method not only circumvents the material waste from heterogeneous wafer bonding techniques, but also improves the heterointerface quality to rival lattice-matched epitaxy[30,31]. This improvement is typically achieved through the enrichment of interfacial oxygen content[27,28,31] or the introduction of an ultrathin oxide (UO) interfacial layer[30,34], complemented by the altered mechanical flexibility of nanomembranes as compared to their bulk counterparts[29]. These factors greatly resolve/circumvent the issues associated with strain-induced defects, thermal expansion mismatches, and surface charge imbalance, which are challenging for conventional heteroepitaxy [4,35] and wafer bonding techniques[36-38].

In this study, leveraging this novel grafting method, we report the synthesis and characterization of a lattice-mismatched single-crystalline GaAs/GeSn-MQW/Ge *n-i-p* heterojunction. Detailed scanning transmission electron microscopy (STEM) studies showed the grafted GaAs/GeSn-MQW heterointerfaces were atomically clean, without any significant atomic intermixing, further confirming the robustness of the grafting approach. Photodiodes were then fabricated from both grafted GaAs/GeSn-MQW/Ge and reference epitaxial Ge/GeSn-MQW/Ge heterojunctions. Compared to conventional epitaxial photodiodes, the tailored electronic band alignment and optical light distribution in the grafted GaAs/GeSn-MQW/Ge heterojunction photodiode result in distinct device behaviors, including an extended detection wavelength range from the visible to near-infrared. Meanwhile, critical performance metrics are significantly improved. The dark current density is at least five orders of magnitude lower than state-of-the-art GeSn photodiodes. Moreover, the photoresponsivity exhibits a remarkable enhancement, achieving approximately a sevenfold increase in the VIS range and a threefold improvement in the NIR range compared to the reference epitaxial photodiode. While this study employs the GaAs/GeSn-MQW/Ge heterostructure to demonstrate the potential for manipulating and enhancing devices, our strategy can be extended to other semiconductor systems, including III-nitrides, oxides, and 2D materials.

**Results**

## I: Design Principle

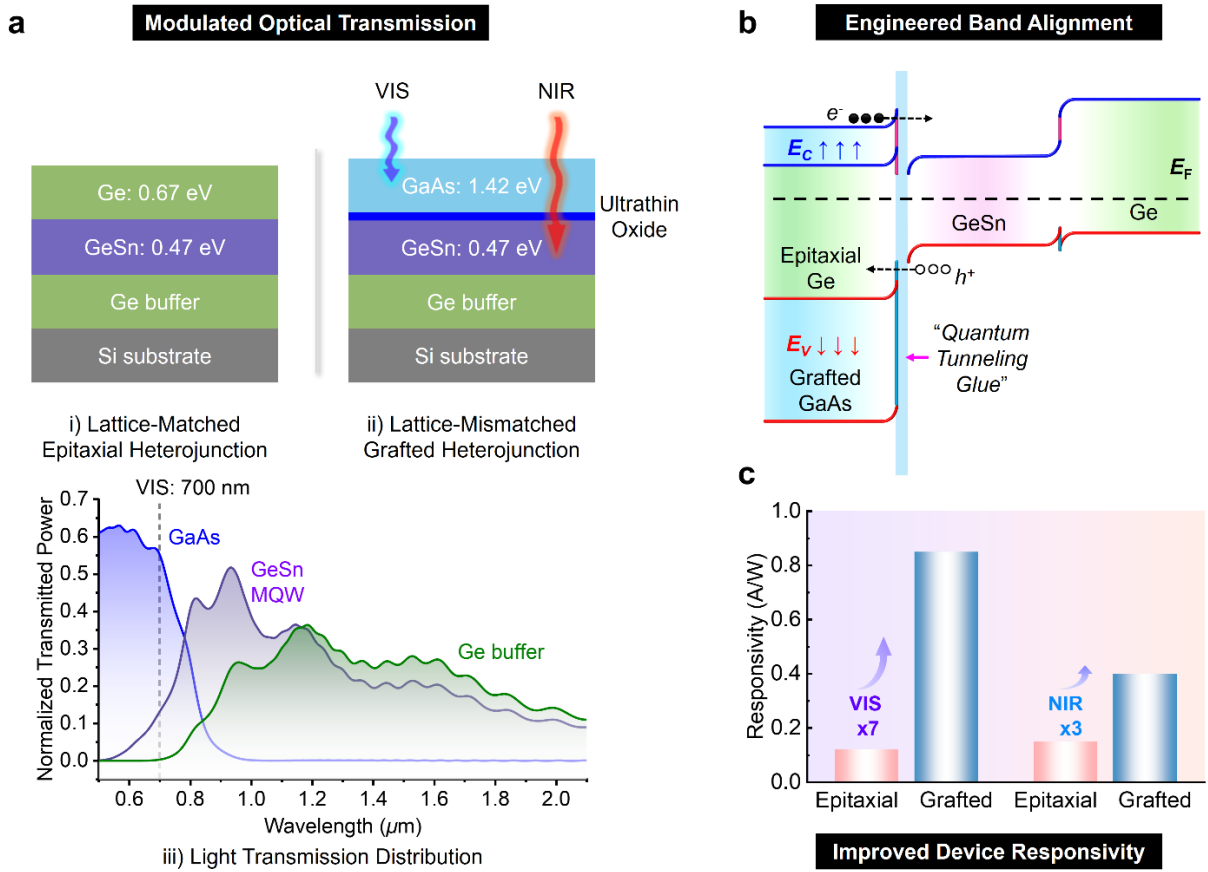

**Fig. 1. Engineered optical field distribution and electronic band alignment.** (a) Schematic comparison of i) conventional lattice-matched epitaxial Ge/GeSn heterojunction and ii) the lattice-mismatched grafted GaAs/GeSn heterojunction, as created in this work. iii) Simulated light transmission profiles show visible-light absorption in the GaAs nanomembrane layer and light penetration into the GeSn/Ge regions at longer near-infrared (NIR) wavelengths. The integration of a GaAs nanomembrane enables modulation of the optical field distribution across the heterostructure. (b) Band alignment at thermal equilibrium for the grafted GaAs/GeSn heterojunctions, superimposed with that of the epitaxial Ge/GeSn reference. The GaAs nanomembrane introduces larger conduction band ($\Delta E_C$) and valence band offsets ($\Delta E_V$), which, combined with quantum tunneling through the ultrathin oxide interlayer, enables effective manipulation of charge carrier transport. (c) Measured photoresponsivity showing the significant performance improvements achieved with the grafted heterojunction in both VIS and NIR ranges, compared to conventional epitaxial devices.

Conventional epitaxial GeSn heterojunctions, as illustrated in **Fig. 1(a i)**, are constrained by the lattice-matching rule, which tolerates only minor lattice mismatches and consequently limits the ability to fine-tune the heterojunction properties. By replacing the original epitaxially grown Ge cap with a foreign GaAs layer, as shown in **Fig. 1(a ii)**, the grafted heterojunction achieves altered optical and electronic characteristics. Specifically, this GaAs layer acts as a highly efficient light-absorbing layer in the VIS range, while longer NIR wavelengths penetrate deeper into the GeSn/Ge regions, as confirmed by the simulated light transmission distribution in each layer in **Fig. 1(a iii)**. Furthermore, the significantly larger bandgap of GaAs compared to GeSn and Ge introduces pronounced conduction band ($\Delta E_C$) and valence band ($\Delta E_V$) offsets at the heterointerface, as depicted in **Fig. 1(b)**. These enlarged band offsets provide greater capacity to manipulate charge carrier transport across the heterojunction, enabling precise control over device operation and performance. This unique engineered band alignment, coupled with optimized optical field distribution, leads to a notable improvement in photoresponsivity from the photodiodes employing the heterojunction designed. As evidenced in **Fig. 1(c)**, the grafted photodiode exhibits superior photo responsivities in both the VIS and NIR ranges, significantly outperforming those measured from reference devices fabricated with the same epitaxial wafer.

## II. Characterization of GaAs/GeSn-MQW/Ge Heterojunction

**Figure 2** illustrates the synthesis and characterization of the grafted GaAs/GeSn-MQW/Ge heterojunction. To create the grafted GaAs/GeSn-MQW/Ge heterojunction, we employed a sequence of steps involving photolithography, dry etching, and transfer printing techniques, as illustrated in **Fig. 2(a i)** to **2(a v)**. This process results in a high-quality GaAs/GeSn-MQW/Ge heterostructure ready for further characterization. A brief description of the procedure can be found in the **Materials and Methods** section.

During the creation of grafted GaAs/GeSn-MQW/Ge heterojunction, we specifically employ the sub-nanometer thin ALD-$Al_2O_3$ as the interlayer between GaAs and GeSn-MQWs due to its uniform and conformal coverage within only a few atomic layers, and its ability to passivate both surfaces of GaAs and GeSn. The detailed ALD-$Al_2O_3$ process description and illustration can be found in **Note S2** and **Fig. S3**. Moreover, the ALD-$Al_2O_3$ interlayer serves as a physical buffer that separates the adjacent GaAs and GeSn-MQW, avoiding direct lattice contact, which would otherwise introduce increased interface states due to their lattice mismatch. More importantly, the ultra-thinness of this ALD-Al$_2$O$_3$ layer allows charge carriers to traverse the heterointerface through quantum tunneling[34].

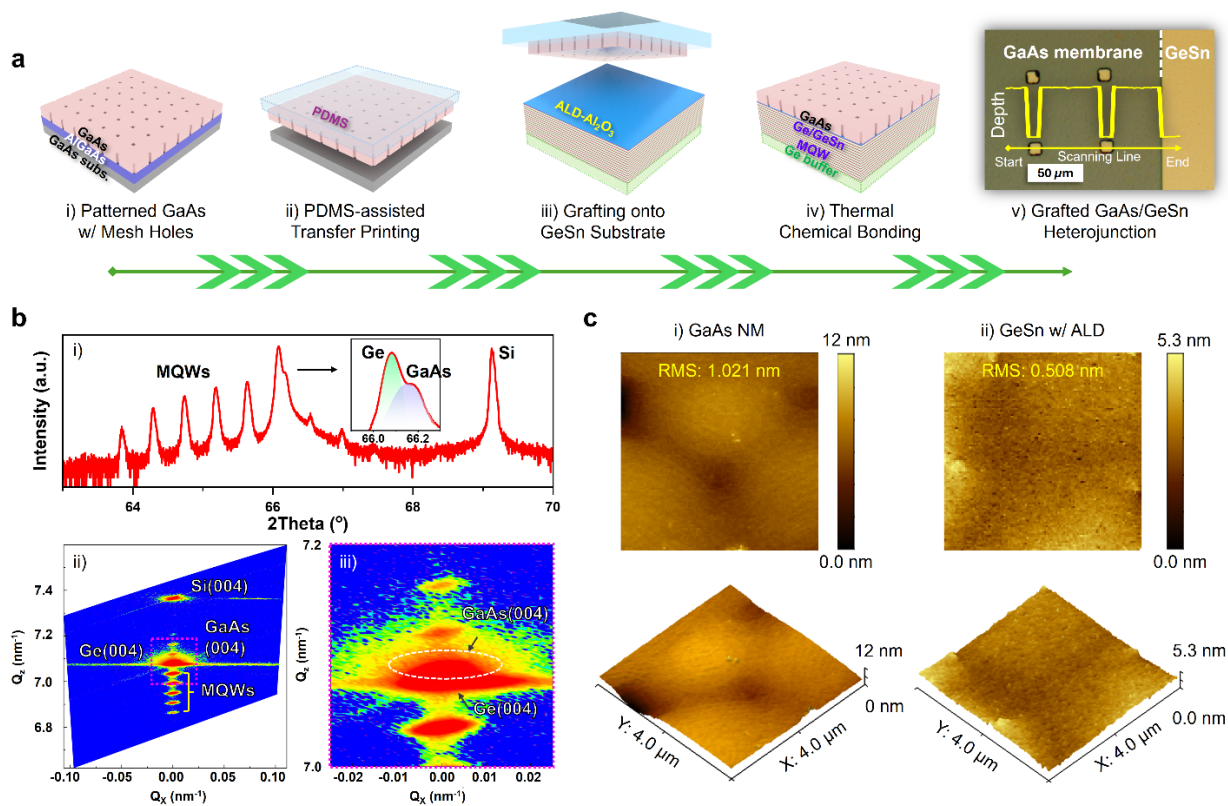

**Fig. 2. Synthesis and characterization of grated GaAs/GeSn-MQW/Ge heterojunction.** (a) The grafting technique involves i) patterning and dry etching of a GaAs/AlGaAs/GaAs epitaxial substrate, ii) sacrificial undercut and release of the GaAs nanomembrane via a PDMS stamp, iii) transfer printing of the GaAs nanomembrane to an ALD-Al$_2$O$_3$-coated GeSn-MQW/Ge substrate, and iv) thermal chemical bonding to form the heterojunction; v) Microscopic image depicting GaAs nanomembrane grafted to GeSn substrate; (b) High-resolution X-ray diffraction (XRD) 2θ-ω spectrum and reciprocal space mapping (RSM) analysis of the grafted GaAs/GeSn-MQW/Ge heterojunction. The 2θ scanning curve is presented in i), the RSM figure is shown in ii), and the high resolution RSM of GaAs/Ge peaks is highlighted in iii); (c) Atomic force microscope (AFM) surface topology mapping of i) the grafted GaAs nanomembrane, and ii) the GeSn-MQW/Ge substrate coated with ALD-Al$_2$O$_3$.

High-resolution X-Ray diffraction (HR-XRD) and reciprocal space mapping (RSM) analysis were performed to characterize the single-crystallinity and strain status of both the grafted GaAs epilayer and the GeSn substrate, as presented in **Fig. 2(b)**. In the 2Theta-Omega (2θ-ω) scanning curve along the (004) orientation (see **Fig. 2(b i)**), distinct peaks for GeSn-MWQs, Ge, GaAs, and Si can be observed from left to right. Notably, the GaAs and Ge peaks are located close to each other in the spectrum due to their similar lattice constants (5.653 Å versus 5.658 Å, respectively). The widths of the GaAs (004) and Ge (004) peaks, fitted with a Gaussian function, are 0.089° and 0.044°, respectively, indicating exceptional crystal quality. From the RSM analysis in **Fig. 2(b ii)**, it is evident that the diffraction peaks of the GeSn-MQWs and Ge buffer are situated at the same in-plane reciprocal lattice vector $Q_x$, indicating that the GeSn-MQWs are pseudomorphically grown and compressively strained to the Ge buffer, with no obvious observable strain relaxation occurring after Ge cap removal and GaAs nanomembrane grafting[21,25,39]. A slight separation at reciprocal lattice vector $Q_z$ of the GaAs and Ge peaks is more clearly observed in the high-resolution RSM image, as shown in **Fig. 2(b iii)**. The observed narrow widths, coupled with the RSM analysis, unambiguously confirm the high-quality epitaxial growth of GeSn-MWQs and grafting of GaAs nanomembranes. The details of XRD and RSM characterization are seen in the **Materials and Methods** section.

**Figures 2(c i)** and **(c ii)** present the surface morphology of the grafted GaAs nanomembrane and the ALD-$Al_2O_3$-passivated GeSn substrate, respectively, as obtained by atomic force microscopy (AFM). Both images are captured over a scanning area of $4 \times 4 \ \mu m^2$. The GeSn substrate, following the Ge cap removal and ALD deposition of a sub-nanometer thick $Al_2O_3$ layer, has a root-mean-square (RMS) roughness of 0.508 nm. Meanwhile, the GaAs nanomembrane, even after undergoing multiple release and transfer steps, maintained its smoothness, with an RMS roughness of 1.021 nm.

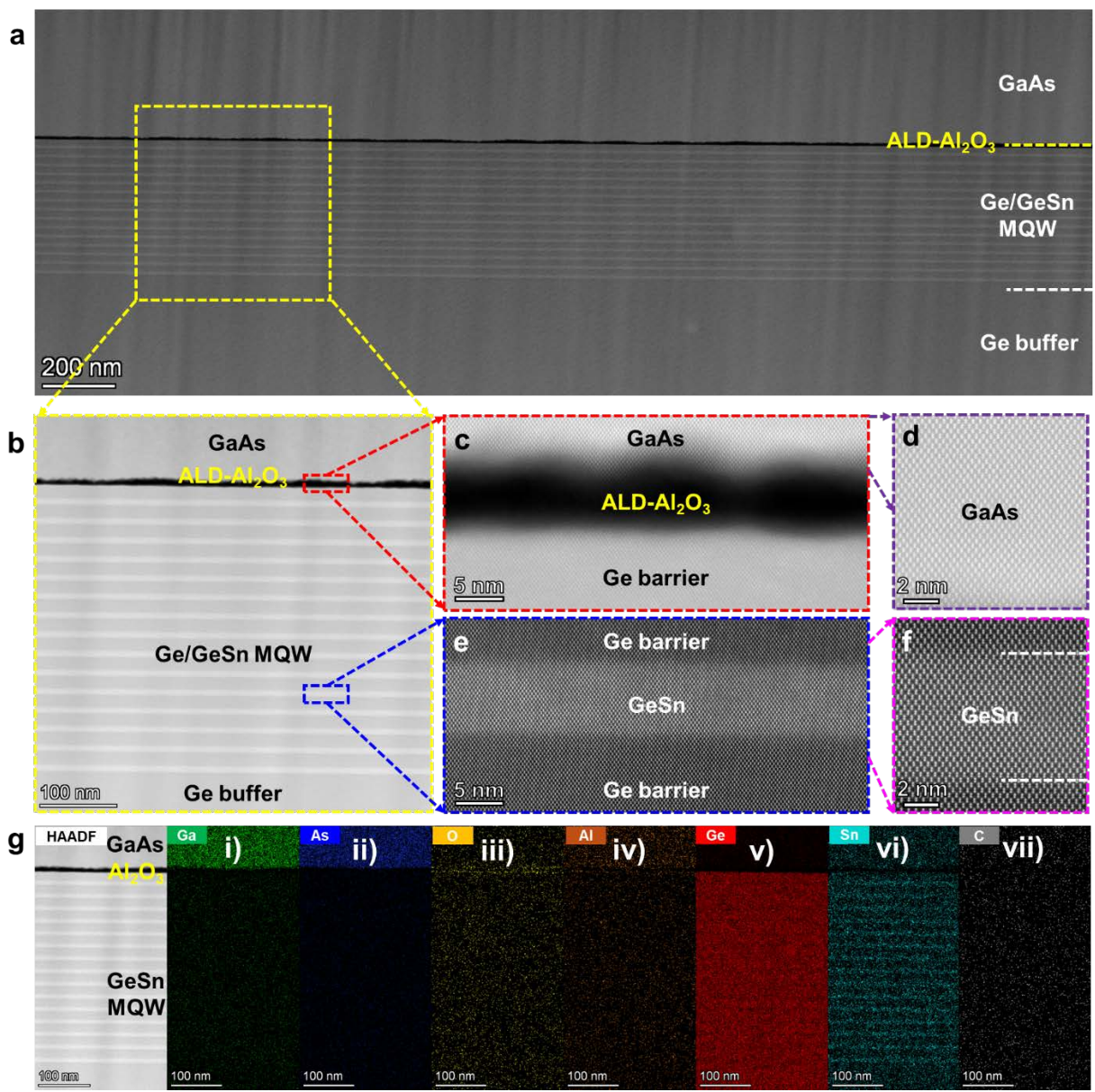

**Fig. 3. STEM characterization.** (a) STEM-HAADF image showing the GaAs/GeSn-MQW-Ge heterojunction structure, demonstrating excellent uniformity and structural integrity over a large scale; (b) A magnified view of the outlined region in (a), showing the precise placement of the GaAs nanomembrane atop the GeSn-MQW region; (c) High-resolution HAADF image taken from the region outlined by the red frame in (b), showing the GaAs/ALD-Al$_2$O$_3$/GeSn-MQW junction; (d) High-resolution HAADF image revealing the preserved single-crystalline nature of the grafted GaAs nanomembrane; (e) Enlarged view of the epitaxially grown Ge/GeSn MQWs of the region outlined by the blue frame, complemented by (f) a high-resolution HAADF image highlighting the high-quality lattice-matched growth within the MQW region; (g) A HAADF

image together with EDS element maps of i) Ga, ii) As, iii) O, iv) Al, v) Ge, vi) Sn, and vii) C, from left to right, illustrating the elemental distributions in the junction.

The structural and interfacial properties were further studied utilizing scanning transmission electron microscopy (STEM). **Figure 3** presents the scanning transmission electron microscopy (STEM) analysis of the grafted GaAs/GeSn-MQW/Ge heterojunction, featuring an interfacial ALD-$Al_2O_3$ layer. As shown in **Fig. 3(a)**, a STEM high-angle annular dark-field (HAADF) image, the original epitaxial Ge cap layer has been completely removed via a precisely controlled dry etching process, ensuring no over-etching of the underlying GeSn-MQW region. The entire heterojunction has excellent uniformity over a micrometer scale, with no observable defects or voids at the grafted heterointerface. A magnified view of the junction is provided in **Fig. 3(b)**, highlighting the grafted GaAs/ALD-$Al_2O_3$/GeSn-MQW heterointerface and the epitaxially grown Ge/GeSn MQW structure. The high-resolution image in **Fig. 3(c)** reveals an amorphous ALD-$Al_2O_3$ interlayer with a thickness of approximately 5 nm. This is wider than the as-deposited thickness of ~0.5 nm, indicating the redistribution of interfacial atoms, primarily oxygen, and the formation of chemical bonds with the adjacent GaAs nanomembrane and Ge barrier. This phenomenon is consistent with our previous reports[30,40,41] and highlights the role of the interlayer in providing effective double-sided interface passivation while facilitating a chemically bonded heterostructure. The preserved single crystallinity of the GaAs nanomembrane after the grafting process is clearly visible in **Fig. 3(d)**. Similarly, the lattice-matched epitaxial growth of the Ge/GeSn MQWs is confirmed in **Fig. 3(f)**, with no observable degradation in lattice quality following the Ge cap removal and grafting processes. This underscores the robustness of our grafting approach. Elemental mapping via STEM spectrum imaging (SI) using X-ray signals further validates the integrity of the grafted heterojunction, as shown in **Fig. 3(g)**. The elemental maps confirm no significant atomic intermixing (*e.g.*, Ga, As, Ge, and Sn) across the heterointerface during the grafting process. The enriched oxygen distribution originating from the ALD-$Al_2O_3$ interlayer at the interface (**Fig. 3(g iii)**) indicates effective passivation on both the GaAs and GeSn-MQW sides. Finally, the negligible carbon content (**Fig. 3(g vii)**) indicates an atomically clean interface, further reinforcing the quality of the grafted heterojunction. A more comprehensive HAADF imaging and elemental mapping of the entire heterostructure can be found in **Fig. S4**. The experimental details of STEM characterization are provided in the **Materials and Methods** section.

The STEM, XRD and AFM characterizations collectively confirm the quality of our membrane grafting process. This process ensured that the GaAs/GeSn-MQW/Ge grafted heterojunction retained both the high crystallinity and smooth surface morphology inheriting from the original epilayers. These results establish a strong basis for subsequent device fabrication and further characterization.

**III: Photoelectrical Characteristics of Grafted GaAs/GeSn-MQW/Ge *n-i-p* Device**

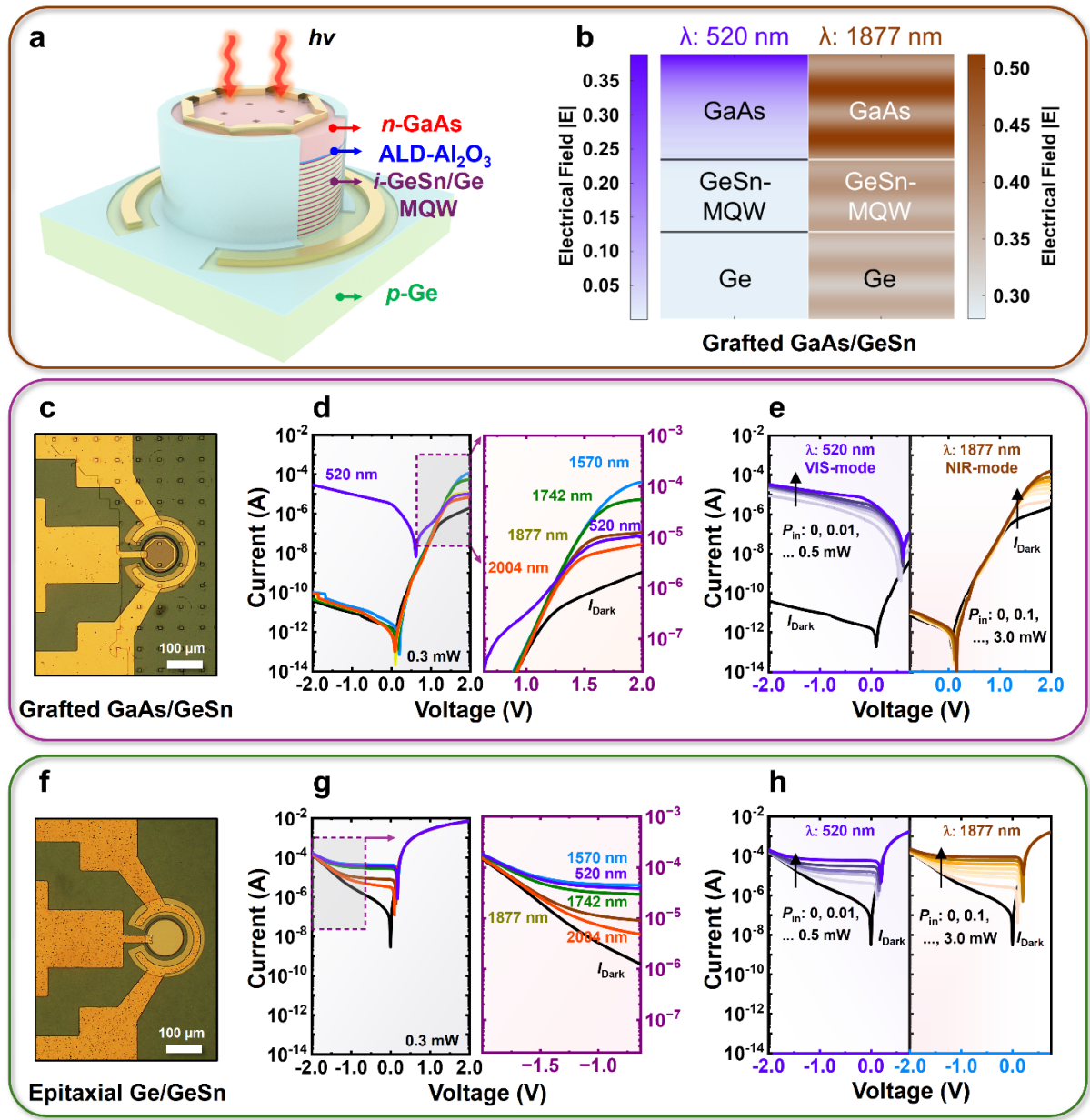

**Fig. 4. Optical and electrical characteristics of grafted photodiodes.** (a) Schematic representation of the completed grafted photodiode. (b) Simulated light field distribution within the grafted GaAs/GeSn-MQW/Ge photodiode, under 520 nm illumination (left panel) and 1877 nm illumination (right panel). The GaAs layer predominantly absorbs light in the visible range, while the GeSn-MQW and Ge buffer regions absorb light in the near-infrared range. (c) Microscopic image of a grafted photodiode. (d) Current-voltage (I-V) characteristics of the grafted photodiode illuminated at various VIS and NIR wavelengths with a fixed power of 0.3 mW. (e) I-V characteristics under different incident power levels, with the left panel illuminated at 520 nm and the right panel fixed at 1877 nm. (f) Microscopic image of a reference epitaxial photodiode. (g) I-V characteristics of a reference epitaxial photodiode under various incident light wavelengths at a fixed power of 0.3 mW. (h) I-V characteristics at different powers and different wavelengths of 520 nm (left) and 1877 nm (right). See **Supplementary Table S1** for detailed incident powers.

After characterizing the material properties of the grafted GaAs/GeSn-MQW/Ge heterojunction, photodiodes were fabricated to showcase the capability to fine-tune device characteristics using this grafted heterojunction, which is otherwise challenging, if not impossible, by conventional heteroepitaxy. The details of this device fabrication are provided in the **Materials and Methods** section.

The schematic of the grafted GaAs/GeSn-MQW/Ge heterojunction photodiode is presented in **Fig. 4(a)**. The simulated optical field within the grafted GaAs/GeSn-MQW/Ge under 520 nm (VIS) and 1877 nm (NIR) illumination is shown in **Fig. 4(b)**. In grafted heterostructure, at 520 nm, most of the incident light is absorbed by the grafted GaAs layer, due to its photon energy (2.38 eV) being much larger than the band gap of GaAs (1.42 eV). At the longer wavelength of 1877 nm (0.66 eV), the incident light penetrates through *n*-GaAs and is absorbed in the *i*-GeSn-MQW and *p*-Ge buffer regions, as quantified in **Fig**. **1(a iii)**. This distribution pattern persists under other NIR illumination wavelengths, including 1570, 1742, and 2004 nm, as shown in **Fig. S12**. Moreover, for comparison, the light field distribution, and transmission/reflectance spectra are also simulated for the epitaxial structure, as seen in **Fig. S11 and Fig. S13**.

Microscope images of the fabricated GaAs/GeSn-MQW/Ge *n-i-p* grafted photodiode and Ge/GeSn-MQW/Ge *n-i-p* reference photodiode, are presented in **Figs. 4(c)** and **4(f)**, respectively, both featuring the same mesa diameter of 100 *μ*m. The current-voltage (I-V) characteristics of these two types of photodiodes under different incident light wavelengths, ranging from visible to near infrared, are illustrated in **Figs. 4(d)** and **4(g)**, respectively. Under dark conditions, both photodiodes present rectifying behavior. The rectification ratio of the grafted photodiode reaches approximately $10^4$ at $\pm 2$ V, nearly two orders of magnitude higher than that of the reference photodiode. This is attributed to the increased band offsets of grafted GaAs/GeSn-MQW heterojunction as compared to those of the Ge/GeSn-MQW case (see also in **Fig. 1(b)**). In particular, the dark current of the grafted photodiode is significantly suppressed, reaching about $10^{-10}$ A at -2 V, which is roughly six orders of magnitude lower than that of the reference photodiode.

Under an incident power of 0.3 mW, the reference sample displays conventional GeSn photodiode behavior operating at the negatively biased region, with the reverse current increasing due to photogenerated carriers driven by the electric field and collected by the electrodes. The photocurrent of the reference photodiode, as calculated from the increment of reverse current from the dark current, varies with different incident wavelengths. The photocurrent increases in the sequence of 2004, 1877, 1742, 520, and 1570 nm, as shown more clearly in the magnified view in the right panel of **Fig. 4(g)**.

By contrast, the grafted photodiode presents a distinct dual-mode operation behavior when illuminated under the same incident wavelengths and power. Under reverse bias, significant photocurrent is observed for 520 nm light, with negligible photocurrent detected for all NIR wavelengths. However, the photo response from these NIR wavelengths can be clearly observed under forward bias, with the photocurrent increasing as the wavelengths decrease (photon energy increases), *i.e.*, 2004, 1877, 1742, and 1570 nm, respectively, as shown more clearly in the magnified view in the right panel of **Fig. 4(d)**. Therefore, the grafted photodiode operates at both VIS and NIR wavelengths, as to be analyzed in detail in the following **Device Physics** section.

Furthermore, the photo response under varying incident power was recorded for both the grafted and reference epitaxial photodiodes, as shown in **Figs. 4(e)** and **4(h)**, respectively. Under reverse bias, as illustrated in the left panel of **Fig. 4(d)**, the photocurrent of the grafted photodiode saturates immediately when illuminated with 520 nm light, even at the minimum incident power of 10 $\mu$W (reaching the limit of our measurement system). This saturation is also reflected in the sharp drop in photoresponsivity observed in **Fig. 6(b)**. In contrast, shown in **Fig. 4(g)**, the reference photodiode exhibits only a moderate response at this visible wavelength, as constrained by the absorption of epitaxial Ge cap layer. Under forward bias, as depicted in the right panel of **Fig. 4(e)**, the photocurrent of the grafted photodiode increases with the incident power at 1877 nm. Similar behavior is observed at other NIR wavelengths (see **Fig. S6**). For the reference photodiode, all photo responses are recorded under reverse bias, including those for 520 nm and 1877 nm wavelengths (**Fig. 4(h)**), as well as for additional NIR wavelengths shown in **Fig. S6**.

## IV: Device Physics

Unlike the reference epitaxial photodiode, the grafted photodiode exhibits a unique dual-mode operation over the VIS-NIR regions. This distinct behavior arises from its engineered optical field distribution and precisely tailored band alignment.

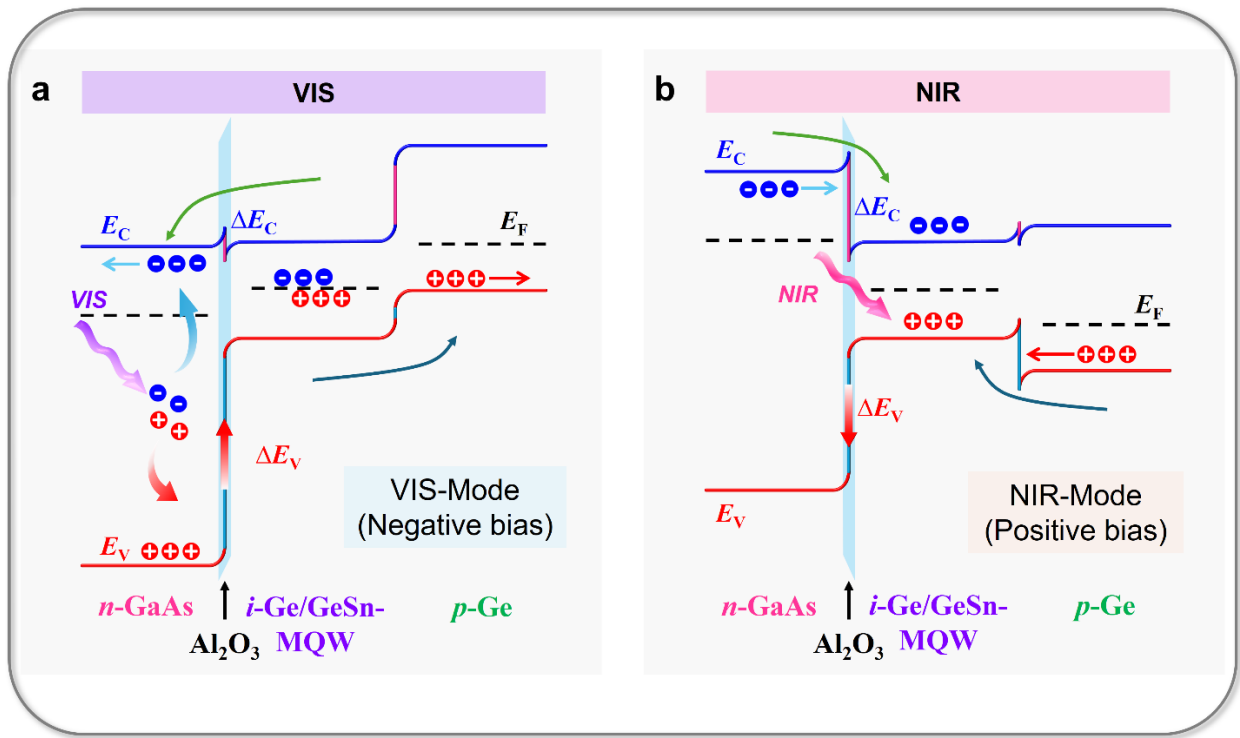

**Fig. 5. Schematic representations of charge transport mechanisms.** (a) Schematic of photoelectric processes and charge transport in the grafted photodiode operating under VIS-mode (negative bias). Light absorption occurs primarily in the GaAs layer, with photogenerated holes driven by the valence band offset ($\Delta E_V$) and electrons being collected under bias; (b) Schematic under NIR-mode (positive bias). Electron injection is greatly facilitated by the large conduction band offset ($\Delta E_C$), while holes overcome the valence band barrier ($\Delta E_V$) through photon excitation and external electric fields.

**Figures 5(a)** and **(b)** schematically represent the band alignment of the grafted photodiode at biased conditions, specifically reverse bias (VIS-mode) and forward bias (NIR-mode). A comprehensive depiction of the photoelectric processes, including a comparison with the reference photodiode, is available in **Fig. S7**.

In reverse bias (VIS-mode) for the grafted photodiode, the incident 520 nm light is predominantly absorbed by the grafted GaAs layer, resulting in photogenerated electron-hole pairs, as illustrated in **Fig. 4(b)** and **Fig. 5(a)**. In the meantime, the increased valence band offset ($\Delta E_V$) between GaAs and GeSn serves as an accelerator, enhancing hole transport and improving the visible light response. Conversely, the elevated conduction band offset ($\Delta E_C$) at the GaAs/GeSn-MQW heterojunction functions as a potential barrier for electrons, particularly at longer NIR wavelengths where photogenerated carriers are primarily shifted to the GeSn-MQW region. Consequently, charge separation and collection are significantly hindered due to this electron barrier (see also **Fig. S7**), as evidenced by suppressed photo response at these NIR wavelengths (see also **Fig. 4(d)** and **Fig. S6**). As a comparison, the conduction and valence band offset of the epitaxial reference Ge/GeSn-MQW heterojunction are substantially smaller (*i.e.*, $\Delta E_C$: 0.089 eV and $\Delta E_V$: 0.116 eV[21]), allowing for effective collection of photogenerated charge carriers from the GeSn-MQW region at NIR wavelengths, as schematically illustrated in **Fig. 1(b)**.

Under the forward bias (NIR-mode), as shown in **Fig. 5(b)**, the large conduction band offset $\Delta E_C$ at the grafted GaAs/GeSn heterointerface facilitates electron injection by a built-in electric field near the junction. Simultaneously, the potential barrier $\Delta E_V$ for the photogenerated holes can be overcome by photon excitation, aided by the externally applied electric field. This results in a substantial increase in injection efficiency, approximately by a factor of $\exp(\Delta E_g/kT)$, where $\Delta E_g$ is the band gap difference between GaAs and GeSn, and $kT$ is the product of the Boltzmann constant and temperature. Consequently, the photocurrent of the grafted photodiode presents a sharp increase beyond a critical voltage (*i.e.,* around +1V in our device). In contrast, the reference epitaxial Ge/GeSn-MQW heterojunction has only modest $\Delta E_V$ and $\Delta E_C$ and does not display such behavior, as seen in **Fig. 4(g)** and **Fig. S6**. Therefore, the unique grafted heterojunction results in device behaviors that are challenging to achieve through conventional epitaxy methods.

**V: Performance Metrics and Benchmarking of GaAs/GeSn-MQW/Ge *n-i-p* Photodiode**

Following the physics analysis, the performance metrics of the grafted photodiode are quantified and presented in **Fig. 6**. As shown in **Fig. 6(a)**, the grafted photodiode exhibits an exceptionally high photo-to-dark current ratio ($I_{ph}/I_{dark}$) of around $10^5$ at the VIS wavelength of 520 nm. This can be attributed to the enhanced light absorption in the grafted GaAs membrane and improved charge carrier transport at the GaAs/GeSn-MQW heterointerface at this wavelength, as analyzed previously in **Figs. 1(a)**, **1(b)** and **Fig. 5(a)**. In the NIR wavelength range, the overall $I_{ph}/I_{dark}$ ratio shows a decreasing trend as the wavelength increases. Moreover, for each individual wavelength, the $I_{ph}/I_{dark}$ ratio increases with the increasing incident power.

The photoresponsivity (A/W), calculated as the ratio of photocurrent ($I_{ph}$) to incident power ($P_{in}$), is depicted in **Fig. 6(b)** across various wavelengths. At 520 nm, an early sharp decrease in responsivity is observed as the photogenerated electron-hole pairs quickly reach saturation. Further increases in incident light power do not significantly contribute to additional photocurrent, resulting in a decrease in photoresponsivity. Similar decreasing trends are observed at longer NIR wavelengths due to similar generation-recombination processes.

**Figure 6(c)** illustrates the photoresponsivity of the grafted photodiode at 520 nm (VIS) and 1570 nm (NIR) as a function of applied bias, highlighting a dual-mode operation with the VIS response occurring mainly at negative bias and the NIR response at positive bias. Moreover, the photoresponsivity in both modes shows a monotonic increase with the absolute value of the applied bias. This trend can be attributed to the fact that higher applied voltage enhances the collection efficiency of photogenerated carriers, thereby increasing photocurrent and photoresponsivity at a fixed incident power. In **Fig. 6(d)**, when the photodiode is operated at NIR mode, the photocurrent $I_{ph}$ decreases as the incident wavelength grows. This phenomenon aligns with previous report[21] and with our reference epitaxial photodiode (**Fig. S8**), where photocurrents are primarily generated from GeSn-MQW and Ge buffer regions under NIR illuminations.

In **Fig. 6(e)**, the photoresponsivity at various wavelengths (520, 1570, 1742, 1877, and 2004 nm) is summarized, demonstrating a broad spectral response range that includes the visible, telecommunication (O- to U-bands), and 2-$\mu$m bands. The highest photoresponsivity is observed at 520 nm, while the lowest is at 2004 nm. This extensive spectral response underscores the versatility and high performance of our grafted photodiode, highlighting its potential for diverse applications across different wavelength ranges.

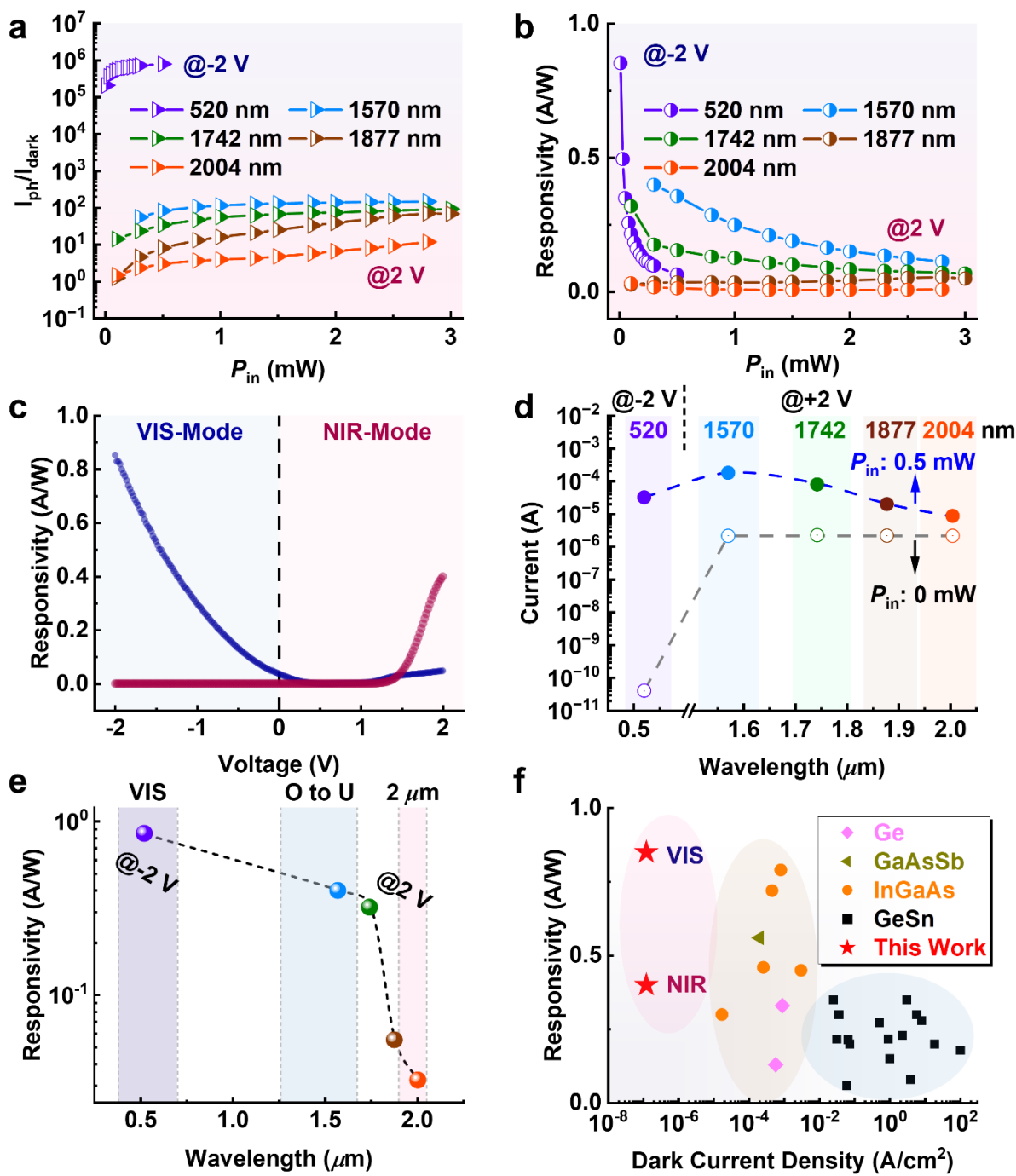

**Fig. 6. Performance metrics and benchmarking of grafted photodiodes.** (a) Photo-to-dark current ratio ($I_{ph}/I_{dark}$) versus incident power ($P_{in}$) of the grafted photodiode for various illumination wavelengths. (b) Photoresponsivity versus incident power ($P_{in}$) of the grafted photodiode under various illumination wavelengths. (c) Photoresponsivity at the VIS wavelength of 520 nm and the NIR wavelength of 1570 nm under varying applied voltage, demonstrating a dual-mode response with the VIS mode operating at negative bias and the NIR mode operating at positive bias.(d) Summarized currents under dark ($P_{in}$ = 0 mW) and illumination ($P_{in}$ = 0.5 mW) conditions, extracted at multiple wavelengths (520, 1570, 1742, 1877, and 2004 nm) and bias voltages (±2 V) (e) Summary of photoresponsivity at various wavelengths: 520, 1570, 1742, 1877, and 2004 nm for ±2 V. (f) Benchmarking of the grafted GaAs/GeSn-MQW/Ge *n-i-p* photodiode. The graph compares the responsivity and dark current density of our work (highlighted with red stars) against previous GeSn photodiodes[11–25], and other material systems, including Ge[42,43], GaAsSb[44], and InGaAs[45–49]. A more detailed metrics table and benchmarking comparison are offered in **Supplementary Tables S2, S3** and **S4**.

In **Fig. 6(f)**, we benchmark the performance of our grafted GaAs/GeSn-MQW/Ge heterojunction photodiode against a range of epitaxial GeSn-based *p-i-n* photodiodes[11–25], as well as photodiodes based on other material systems[42–49], using reverse dark current density and photoresponsivity as key metrics. As illustrated, most epitaxial GeSn photodiodes show relatively high dark current density levels, ranging from $10^{-2}$ A/cm$^2$ to $10^2$ A/cm$^2$, while photodiodes from other material systems report lower values, between $10^{-5}$ A/cm$^2$ to $10^{-3}$ A/cm$^2$. Furthermore, the photoresponsivity of previous GeSn devices, measured near 1550 nm, is generally below 0.4 A/W. In stark contrast, our grafted photodiode demonstrates a record-low dark current density of $1.22 \times 10^{-7}$ A/cm$^2$, along with high photo responsivities of 0.85 A/W and 0.40 A/W at measured at 520 nm (VIS) and 1570 nm (NIR), respectively.

**Discussion**

The exceptional performance metrics of the fabricated GaAs/GeSn-MQW/Ge *n-i-p* grafted photodiode in both VIS and NIR ranges underscore the superiority of our semiconductor grafting method. This method allows for strategic engineering of band alignment and optical absorption within the heterostructure created. By incorporating an ultrathin ALD-Al$_2$O$_3$ interlayer for effective interfacial passivation and carefully controlling the thermal budget throughout the process, the grafting technique facilitates efficient charge carrier transport across the heterointerface while minimizing impedance from interface traps.

In conclusion, we demonstrated a significant improvement of photodiode performance via creating a heterojunction composed of valence- and lattice-mismatched, single-crystalline GaAs and GeSn semiconductors. By strategically tailoring the band alignment and light distribution, the fabricated grafted photodiodes achieved a record-low dark-current density of $1.22 \times 10^{-7}$ $A/cm^2$, alongside an extended dual-band response in the visible and near-infrared regions. The devices exhibited high photoresponsivity, with $R_{VIS}$ reaching 0.85 A/W and $R_{NIR}$ reaching 0.40 A/W. This work establishes *semiconductor grafting* as an effective approach for developing novel GeSn-based optoelectronic heterojunction devices, delivering advanced functionalities and superior performance metrics. Furthermore, it provides valuable insights into addressing critical challenges associated with band alignment engineering and interface quality optimization in lattice-mismatched heterostructures. These findings open up new avenues for advancing heterogeneous integration technologies, enabling multifunctional optoelectronic platforms based on diverse material systems, such as III-nitrides, 2D materials, and oxides.

## Materials and Methods

### Materials

The epitaxial structure of GeSn epi consists of ~200 nm $n^+$ Ge cap, 15 pairs of 20 nm/7.5 nm $Ge/Ge_{0.08}Sn_{0.92}$ MQWs, and a ~2 µm $p^+$ Ge buffer layer, all grown on a 12-inch Si substrate. The GaAs epi consists of a 100/500 nm $n^+/n^-$ cap GaAs layer and a 400 nm unintentionally doped (UID) $Al_{0.95}Ga_{0.05}As$ layer, both grown on a 2-inch semi-insulating (SI) GaAs substrate. Detailed epitaxial structures and characterization of the source GaAs and GeSn epi-wafers can be found in Supplementary **Fig. S1**.

### Fabrication of GaAs/GeSn-MQW/Ge heterojunction

Initially, single-crystalline, freestanding GaAs nanomembranes were obtained from source GaAs epitaxial wafers using photolithographic patterning for creating mesh holes and dry etching processes to transfer the pattern into GaAs and expose sacrificial AlGaAs, as seen in **Fig. 2(a i)**. The GaAs nanomembrane was then released and transferred to the ALD-$Al_2O_3$-passivated GeSn substrate using a polydimethylsiloxane (PDMS) stamp, as depicted in **Fig. 2(a ii)** and **Fig. 2(a iii)**, respectively. Finally, a rapid thermal annealing was performed to achieve chemical bonding between ALD-$Al_2O_3$ and the adjacent GaAs and GeSn-MQW, as presented in **Fig. 2(a iv)**. Moreover, comprehensive procedural details of grafting (**Supplementary Note S1**), and optical profiling on grafted GaAs nanomembrane (**Fig. S2**) can be found in **Supplementary Information**.

**Fabrication of Photodiodes**

The fabrication of the grafted GaAs/GeSn-MQW/Ge heterojunction photodiode involves several critical steps. Starting from the synthesized GaAs/GeSn-MQW/Ge heterostructure, the process begins with the formation of a mesa structure using standard photolithography and subsequent dry etching to remove the extra bonded *n*-GaAs layer and expose the *i*-GeSn-MQWs and *p*-Ge buffer. Surface passivation and via openings for electrodes are then created by depositing $SiO_x$ using PECVD and etching through the oxide layer. The fabrication is completed with metallization on the *n*-GaAs and *p*-Ge layers. A detailed description and schematic illustration of the fabrication procedures can be, respectively, found in **Supplementary Note S3 and Fig. S5**.

A schematic of the completed grafted photodiode is provided in **Fig. 4(a)**, highlighting a multilayered device heterostructure, which includes the grafted *n*-type GaAs nanomembrane, the interfacial ALD-$Al_2O_3$, intrinsic GeSn-MQWs, and the underlying *p*-type Ge buffer layer. A microscope image of the grafted device, featuring a 100 $\mu$m inner diameter of mesa, is shown in **Fig. 4(c)**. For a direct comparison with the grafted photodiodes, a reference Ge/GeSn-MQW/Ge *n-i-p* photodiode with the same device footprint was fabricated from the same as-grown epitaxial sample, without removing the *n*-Ge cap. The fabrication for the grafted and reference samples were carried out simultaneously to minimize potential device-performance variations associated with the processing steps. Detailed thermal budget management, which is crucial to structural integrity and device performance, is elaborated in **Supplementary Note S4** and **Table S5.**

**Characterizations**

**XRD characterization**

The XRD 2θ-ω spectrum and RSM mapping was obtained by a Malvern Panalytical Empyrean X-ray diffractometer with a Cu K-α X-ray source. The detailed settings were same as our previous work[30].

**STEM specimen preparation and instrumentations**

A Thermo-Fisher Helios 650 Xe plasma dual-beam focused (PFIB) system was used for electron microscopy specimen preparation. A 30 keV ion beam with beam currents between 4 nA and 30 pA was used for lift-out cutting and initial thin foil thinning. For final thinning and polishing of the thin foils, a 5keV beam with beam current of 30 pA was used.

A Themo-Fisher Spectra 300 transmission electron microscope with a probe corrector was used for characterizing the microstructures of the material. The microscope was operated at 300 keV in STEM mode that gives a spatial resolution of ~70 pm. High-angle annular dark-field imaging and STEM spectrum imaging (SI) using X-ray signals (X-ray energy dispersive spectroscopy, EDS) were performed.

**Data availability**

The data supporting this study's findings are available from the corresponding author upon reasonable request.

**Acknowledgements**

The work is supported by the Air Force Office of Scientific Research (AFOSR) under grant # FA9550-19-1-0102.